\documentclass{desyproc}

\begin{document}
\title{Exclusive photoproduction of $J/\psi$ and $\psi(2S)$ states in $pp$ and $AA$ collisions at the LHC}

\author{{\slshape Magno V. T. Machado}\\[1ex]
High Energy Physics Phenomenology Group, GFPAE  IF-UFRGS \\
Caixa Postal 15051, CEP 91501-970, Porto Alegre, RS, Brazil}

\contribID{machado\_magno}


\acronym{EDS'13} 

\maketitle

\begin{abstract}
In this contribution we report on the investigations of the exclusive production of $J/\psi$ and $\psi (2S)$ states in proton-proton and nucleus-nucleus reactions at the LHC using the theoretical framework of  light-cone dipole formalism.
\end{abstract}

\section{Introduction}

The exclusive vector meson photoproduction has being investigated both experimentally and theoretically in recent years as it allows to test perturbative Quantum Chromodynamics. The quarkonium masses, $m_V$, give a perturbative scale for the problem even in the photoproduction limit.  An important feature of these processes at the high energy regime is the possibility to investigate the Pomeron exchange. For this energy domain hadrons and photons can be considered as color dipoles in the mixed light cone representation, where their transverse size can be considered frozen during the interaction. Therefore, the scattering process is characterized by the color dipole cross section describing the interaction of those color dipoles with the nucleon/nucleus target. Here, we summarize the results presented in Refs. \cite{GGM} where the exclusive production of $J/\psi$ and the radially excited $\psi (2S)$ mesons were studied in $pp$ and $AA$ collisions  in the LHC energy range.  The theoretical framework considered is the light-cone dipole formalism, where the $c\bar{c}$ fluctuation (color dipole) of the incoming quasi-real photon interacts with the nucleon or nucleus target via the dipole cross section and the result is projected in the wavefunction of the observed hadron. At high energies, the transition of the regime described by the linear dynamics of emissions chain to a new regime where the physical process of recombination of partons becomes important is expected. It is characterized by the limitation on the maximum phase-space parton density that can be reached in the hadron wavefunction, the  so-called parton saturation phenomenon. The transition is set by saturation scale $Q_{\mathrm{sat}}\propto x^{\lambda}$, which is enhanced in the nuclear case. The predictions for  $\psi(2S)$ are somewhat new as most part of predictions in literature concern only to the $\psi(1S)$ state.

\section{Theoretical framework and results}
\label{coerente}

 The exclusive meson photoproduction in hadron-hadron collisions can be factorized in terms of the equivalent flux of photons of the hadron projectile and photon-target production cross section \cite{upcs}. The photon energy spectrum, $dN_{\gamma}^p/d\omega$, which depends on the photon energy $\omega$,  is well known \cite{upcs}. The rapidity distribution $y$ for charmonium  photoproduction in  $pp$ collisions can be  written down as,
\begin{eqnarray}
\frac{d\sigma}{dy}(pp \rightarrow   p\otimes \psi \otimes p) & = & S_{\text{gap}}^2 \left[\omega \frac{dN_{\gamma}^p}{d\omega }\sigma (\gamma p \rightarrow \psi(nS) +p)+ \left(y\rightarrow -y \right) \right]. 
\label{dsigdy}
\end{eqnarray}
 The produced state with mass $m_V$ has rapidity $y\simeq \ln (2\omega/m_V)$ and the square of the $\gamma p$ centre-of-mass energy is given by $W_{\gamma p}^2\simeq 2\omega\sqrt{s}$. The absorptive corrections due to spectator interactions between the two hadrons are represented by the factor $S_{\text{gap}}$. The photon-Pomeron interaction will be described within the light-cone dipole frame, where thee probing
projectile fluctuates into a
quark-antiquark pair with transverse separation
$r$ (and momentum fraction $z$) long after the interaction, which then
scatters off the hadron. The cross section for exclusive photoproduction of charmonia off a nucleon target is given by,
\begin{eqnarray}
\sigma (\gamma p\rightarrow \psi(nS) +p) = \frac{1}{16\pi B_V} \left|\sum_{h, \bar{h}} \int dz\, d^2r \,\Psi^\gamma_{h, \bar{h}}\sigma_{dip}(x,r)\, \Psi^{V*}_{h, \bar{h}}  \right|^2,
\label{sigmatot}
\end{eqnarray}
where $\Psi^{\gamma}$ and $\Psi^{V}$ are the light-cone wavefunction  of the photon  and of the  vector meson, respectively.  The Bjorken variable is denoted by $x$, the dipole cross section by  $\sigma_{dip}(x,r)$ and the  diffractive slope parameter by $B_V$.  Here, we consider the energy dependence of the slope using the Regge motivated expression (see \cite{GGM} for details). Similarly, the rapidity distribution $y$ in nucleus-nucleus collisions has the same factorized form,
\begin{eqnarray}
\frac{d\sigma}{dy} (A A \rightarrow   A\otimes \psi(nS) \otimes Y) = \left[ \omega \frac{dN_{\gamma}^A}{d\omega }\,\sigma(\gamma A \rightarrow \psi(nS) +Y )+ \left(y\rightarrow -y \right) \right],
\label{dsigdyA}
\end{eqnarray}
where the photon flux in nucleus is denoted by $dN_{\gamma}^A/d\omega$ and $Y=A$ (coherent case) or $Y=A^*$ (incoherent case). The exclusive photoproduction off nuclei for coherent and incoherent processes can be simply computed in high energies where the large coherence length $l_c\gg R_A$ is fairly valid. the expressions for both cases are given by \cite{Boris},
\begin{eqnarray}
\sigma (\gamma A \rightarrow \psi(nS) +A ) & = & \int d^2b\, |\langle \Psi^V|1-\exp\left[-\frac{1}{2}R_G\sigma_{dip}(x,r) T_A(b)  \right]|\Psi^{\gamma}\rangle |^2, \label{eq:coher} \nonumber \\
\sigma (\gamma A \rightarrow \psi(nS) +A^* )  & = & \frac{1}{16\pi\,B_V(s)}\int d^2b\,T_A(b) |\langle \Psi^V|R_G\sigma_{dip}(x,r) \exp\left[-\frac{1}{2}R_G\sigma_{dip}(x,r)T_A(b)  \right]|\Psi^{\gamma}\rangle|^2, \nonumber
\label{eq:incoh}
\end{eqnarray} 
where $T_A(b)= \int dz\rho_A(b,z)$  is the nuclear thickness function. In the numerical evaluations, we have considered the boosted gaussian wavefunction and the  phenomenological saturation model proposed in Ref. \cite{IIM} (CGC model) which encodes the main properties of the saturation approaches.  The nuclear ratio for the gluon density is denoted by $R_G(x,Q^2=m_V^2/4)$.

First, we compare the study to the data measured by LHCb Collaboration at 7  TeV in $pp$ collisions at the forward region $2.0<\eta_{\pm} <4.5$ \cite{LHCb}, which cover values of Bjorken-$x$ variable down to $x\approx 5\times 10^{-6}$. We assume for the absorption factor the average value $S_{\text{gap}}^2=0.8$.  We obtain $\sigma (pp\rightarrow p+J/\psi+p) \times \mathrm{Br}=698$ pb for rapidity between 2 and 4.5. In terms of muon pseudorapidities we get $\sigma_{pp\rightarrow J/\psi(\rightarrow \mu^+\mu^-)}(2.0<\eta_{\mu^\pm}<4.5) = 298$ pb. This is in good agreement to the experimental result $307\pm 42$ pb \cite{LHCb}. For the $\psi(2S)$ mesons it is obtained $\sigma (pp\rightarrow p+\psi (2S)+p) \times \mathrm{Br}=18$ pb for rapidities $2.0<y<4.5$. Accordingly, we now predict $\sigma_{pp\rightarrow \psi(2S)(\rightarrow \mu^+\mu^-)}(2.0<\eta_{\mu^\pm}<4.5) = 7.7$ pb compared to  $7.8\pm 1.6$ pb measured value \cite{LHCb}. We performed predictions for the next LHC runs in $pp$ mode. We have found $\frac{d\sigma_{J/\psi}}{dy} = 6.2$ nb and  7.9 nb for central rapidities at energies of 8 and 14 TeV, respectively. For the $\psi (2S)$ state the extrapolation gives $\frac{d\sigma_{\psi(2S)}}{dy} = 1.0$ nb and 1.4 nb for the same energies at central rapidity.

\begin{figure}[t]
\centerline{\includegraphics[width=0.4\textwidth]{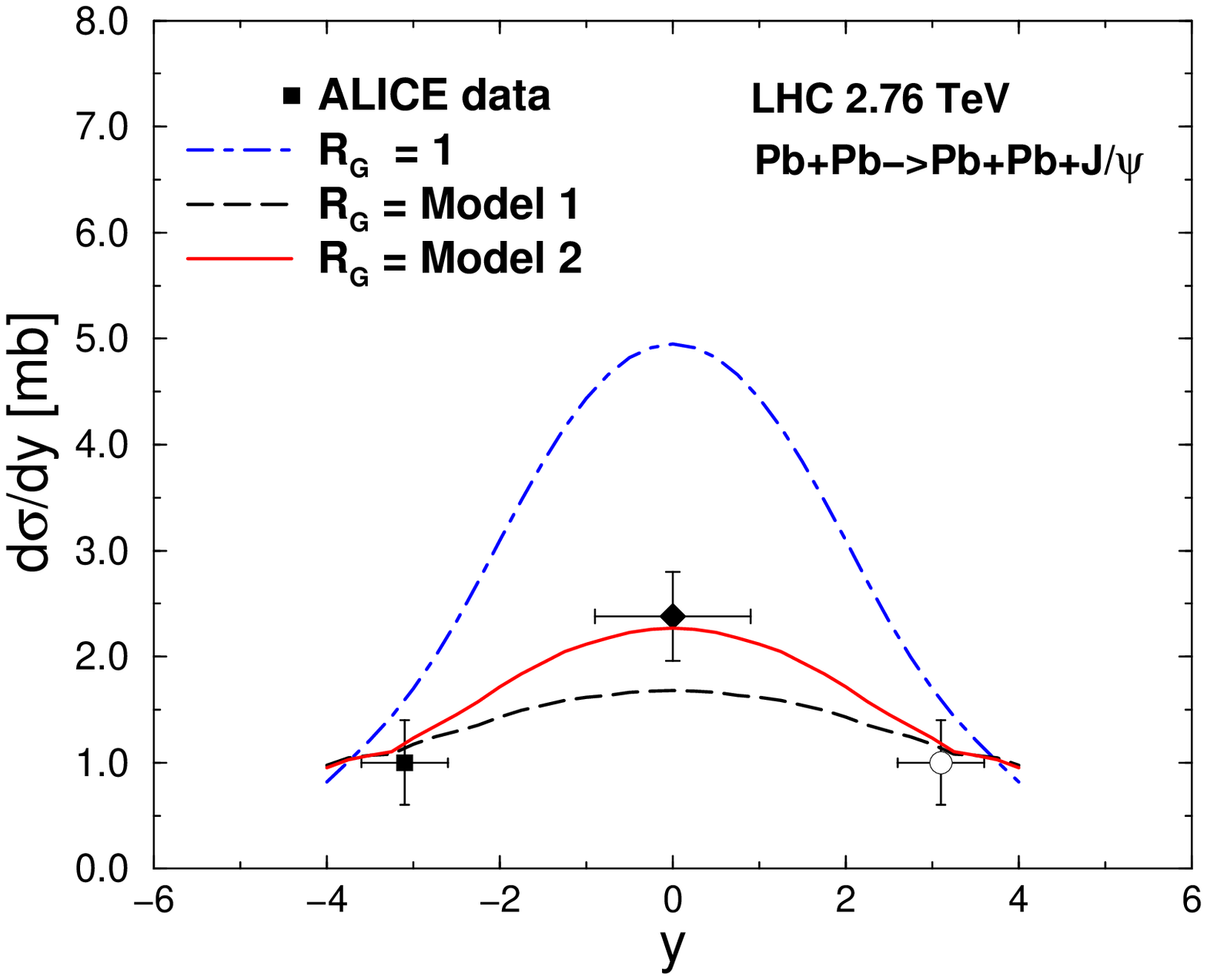}\\\includegraphics[width=0.4\textwidth]{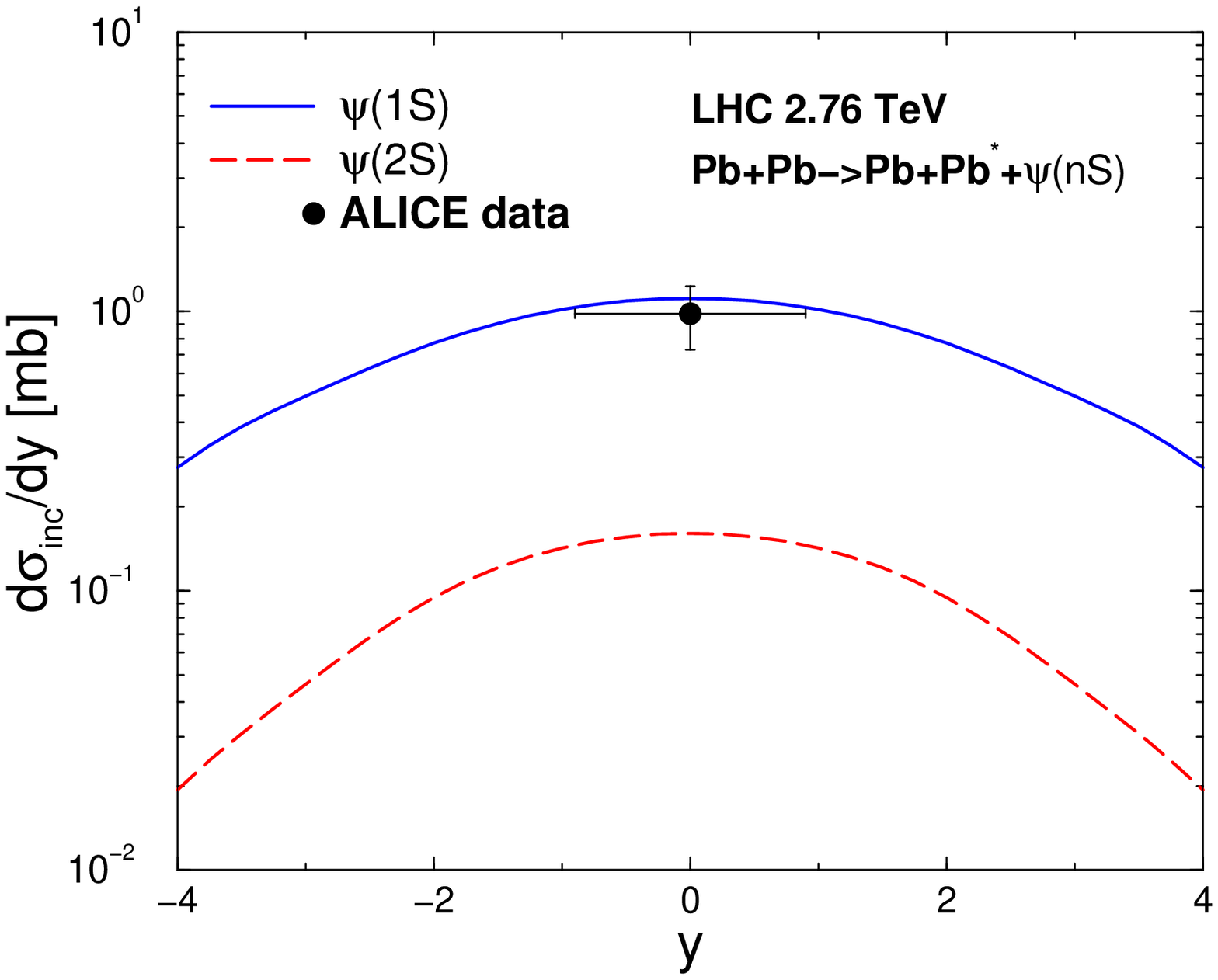}}
\caption{The rapidity distribution of coherent (left panel) and incoherent (right panel) $\psi$(1S,2S) photoproduction at $\sqrt{s}=2.76$ TeV in PbPb collisions at the LHC (see text).}\label{fig:1}
\end{figure}

Finally, in Fig. \ref{fig:1} (left panel) we present the calculations for the rapidity distribution of coherent $\psi(1S)$ state, using distinct scenarios for the nuclear gluon shadowing. The dot-dashed curve represents the result using $R_G=1$. It overestimates the ALICE data \cite{ALICE} on the backward (forward) and mainly in central rapidities.  The situation is improved if we consider nuclear shadowing renormalising the dipole cross section. The reason is that the gluon density in nuclei at small Bjorken $x$ is expected to be suppressed compared to a free nucleon due to interferences. For $R_G$, we have considered the theoretical evaluation of Ref. \cite{GFM}. As a prediction at central rapidity, one obtains $\frac{d\sigma}{dy} (y=0) = 4.95, \,1.68$ and $2.27$ mb for calculation using $R_G=1$, Model 1 (strong shadowing) and Model 2 (weak shadowing), respectively. In Fig. 1 (right panel) is presented the incoherent cross section for both $\psi(1S)$ and $\psi(2)$ states with $R_G=1$.

\section{Acknowledgments}
This work was  partially financed by the Brazilian funding
agency CNPq and by the French-Brazilian scientific cooperation project CAPES-COFECUB 744/12. 
 

\begin{footnotesize}

\end{footnotesize}
\end{document}